\documentstyle[preprint,aps]{revtex}
\tighten
\begin{document}
\draft
\title
{
Langevin equations for continuous time
L\'{e}vy flights
}
\author
{
Hans C. Fogedby
}
\address
{
Institute of Physics and Astronomy\\
University of Aarhus\\
8000 Aarhus C\\
Denmark\\
E-mail: fogedby@dfi.aau.dk
}
\date{\today}
\maketitle
\begin{abstract}
We consider the combined effects of a power law L\'{e}vy step distribution
characterized by the step index $f$ and a power law waiting time
distribution characterized by the time index $g$ on the long time
behavior of a random walker. The main point of our analysis
is a formulation in terms
of coupled Langevin equations which allows in a natural way for the
inclusion of
external force fields. In the anomalous case for $f<2$ and $g<1$
the dynamic exponent $z$ locks onto the ratio $f/g$. Drawing on
recent results on L\'{e}vy flights in the presence of a random
force field we also find that this result is
{\em independent}
of the presence of weak quenched disorder. For $d$ below the critical
dimension $d_c=2f-2$ the disorder is {\em relevant}, corresponding
to a non trivial fixed point
for the force correlation function.
\end{abstract}
\section{Introduction}
Anomalous diffusion arising from non trivial waiting time distributions,
so-called {\ continuous time random walk (CTRW), has been used to model
a variety of physical phenomena \cite{bouch1}. For example, the
dynamics of carrier diffusion and recombination in disordered
media has been described in terms of a CTRW
\cite{scher1,scher2,shle1,klaf1,blumen1}. Anomalous diffusion in an
intermittent dynamical system \cite{man,gei} and in a linear
array of convection cells \cite{pom,car} has also been analyzed
in terms of CTRW.

Anomalous diffusion is also associated with power law step size
distributions, the so-called
L\'{e}vy distributions
\cite{fogedby1,feller,klaf2,blumen2}.
The L\'{e}vy flights (LF) generated by the L\'{e}vy step
distribution have been used to model a variety of physical
processes such as self diffusion in micelle systems \cite{ott}
and transport in heterogeneous rocks \cite{klaf2}.

In the present paper we consider the combined effect of a
power law waiting time distribution and a power law
L\'{e}vy step size distribution, i.e., the case of continuous
time L\'{e}vy flights (CTLF). Since L\'{e}vy flights lead to
superdiffusive behaviour, whereas a power law waiting time
distribution entails subdiffusive behaviour, the combination
of the two yields an interesting description of anomalous
diffusion in the general case \cite{bouch1}.

The key issue in the paper is the discussion of CTLF in
terms of coupled Langevin equations for the position and
the time. This approach is achieved by introducing as an
intermediate variable the path or arc length along a
particular trajectory. Besides the clearer physical
interpretation afforded by Langevin equations as stochastic
equations of motion such an approach also allows in a
natural way for the inclusion of external fields, for
example a drift force field.

In section II we discuss the Langevin equation associated
with power law step and waiting time distributions.
Section III is devoted to the discussion of the associated
Fokker-Planck equations. In section IV we discuss the
probability distribution for the position of the walker as a function of
time. In section V we carry out a scaling analysis in the
absence of an external force field. In section VI we give
a discussion and comment briefly on the role of an external
quenched force field, drawing on recent results.

\section{Langevin equations}

Let us denote the step size distribution by
$\pi({\mbox{\boldmath$\eta$}})$, where ${\mbox{\boldmath$\eta$}}$ is the
elementary microscopic step.
We assume that $\pi({\mbox{\boldmath$\eta$}})$ is normalized, i.e.,
$\int\pi({\mbox{\boldmath$\eta$}})d^d\eta=1$. Likewise, we denote the waiting
time
distribution by $w(\tau)$, where $w(\tau)$ is the probability
of the walker of waiting the time interval $\tau$ at a given
position before performing the next step. Due to causality
$\tau>0$ and the normalization of $w(\tau)$ reads
$\int^\infty_0 w(\tau)d\tau=1$.

Parametrizing the random walk \cite{rw} in the continuum limit
by means of the path parameter or arc length $s$ along the trajectory
we have for the position
of the walker, ${\bf{r}}(s)$, after s `steps',
\begin{eqnarray}
{\bf{r}}(s)=\int^s_0{\mbox{\boldmath$\eta$}}(s')ds',
\end{eqnarray}
or the Langevin equation,
\begin{eqnarray}
\frac{d{\bf{r}}}{ds}={\mbox{\boldmath$\eta$}}(s).
\end{eqnarray}
In Fig. 1 we have shown the parametrization of a particular
trajectory.
Equation (2) is readily generalized in the presence of an external
drift force field ${\bf{F}}$ depending on the position of the walker.
We obtain in this case the Langevin equation,
\begin{eqnarray}
\frac{d{\bf{r}}}{ds}={\bf{F}}({\bf{r}})+{\mbox{\boldmath$\eta$}}(s).
\end{eqnarray}
Similarly, the total elapsed time after s `steps' in the
continuum limit is
\begin{eqnarray}
t(s)=\int_0^s\tau(s')ds',
\end{eqnarray}
implying the Langevin equation,
\begin{eqnarray}
\frac{dt}{ds}=\tau(s).
\end{eqnarray}
The coupled Langevin equations (2), (3), and (5) constitute the
present formulation of continuous time random walk. We are here
considering the separable case where the step distribution
$\pi({\mbox{\boldmath$\eta$}})$ and the waiting time distribution $w(\tau)$ are
statistically independent \cite{bouch1,shle2,weiss}.

In the case of a sharp waiting time distribution
$w(\tau)=\delta(\tau-\tau_0)$, corresponding to a fixed
hopping rate $1/\tau_0$, Eq. (5) becomes deterministic and
can be solved for $s$, i.e., $t=\tau_0s$, and we obtain by
insertion the usual Langevin equation,
\begin{eqnarray}
\frac{d{\bf{r}}}{dt}=\frac{1}{\tau_0}{\bf{F}}({\bf{r}})+\frac{1}{\tau_0}{\mbox{\boldmath$\eta$}}(t).
\end{eqnarray}
In the general case of a non trivial waiting time distribution
we must, however, discuss the coupled equations.

We shall now focus on power law distributions for the waiting
times and steps. For the step ${\mbox{\boldmath$\eta$}}$ we assume an instantly
(in terms of $s$) correlated power law
distribution,
\begin{eqnarray}
\pi({\mbox{\boldmath$\eta$}})d^d\eta\propto\eta^{-1-f}d\eta.
\end{eqnarray}
We assume an isotropic form characterized by the step
index $f$. In order to ensure normalizability we have introduced a
lower cut $\eta\sim a$ of the order of a microscopic length
$a$ and chosen $f>0$. For $f>2$ the second moment,
$\langle\eta^2\rangle=\int\pi({\mbox{\boldmath$\eta$}})\eta^2d^d\eta$,
is finite and a
characteristic step size is given by the root mean square
deviation $\sqrt{\langle\eta^2\rangle}$. For $1<f<2$ the second moment
diverges but the mean step, $\langle\eta\rangle$, is finite. In the
interval $0<f<1$ the first moment diverges and even a mean
step size is not defined \cite{fogedby1}.
In a similar way we assume for the waiting time
$\tau$ the power law distribution,
\begin{eqnarray}
w(\tau)d\tau\propto\tau^{-1-g}d\tau.
\end{eqnarray}
Owing to causality $\tau>0$. In order to guarantee normalizability
we introduce a short time cutoff of order a microscopic time
scale and choose $g>0$. The distribution is characterized by the
time index $g$. For $g>1$ the first moment,
$\langle\tau\rangle=\int w(\tau)\tau d\tau$, is finite, setting a well
defined hopping rate $1/\langle\tau\rangle$. For $0<g<1$
the mean value $\langle\tau\rangle$
diverges and we cannot define a characteristic rate or time
scale.

In order to eventually discuss the anomalous diffusive
characteristics of a random walk driven by the power law noises
${\mbox{\boldmath$\eta$}}$ and $\tau$ we must determine the distribution
function
$P({\bf{r}},t)$ and, in particular, the mean square displacement
\begin{eqnarray}
\langle r^2(t)\rangle\propto Dt^{2/z},
\end{eqnarray}
where $D$ is the diffusion coefficient and $z$ the dynamic
exponent.
Since the random walk takes place in physical time
$t$ we are thus faced with the issue of eliminating the
auxiliary path variable $s$ labelling the walk. Owing to the
stochastic nature of the
Langevin equation (5) for $t$ it is not possible to invert it and
solve it for $s$ and we shall
instead turn to the associated deterministic Fokker-Planck
equations.

\section{Fokker-Planck equations}
Limiting first the discussion to the force free case, i.e.,
${\bf{F}}={\bf{0}}$, the probability distributions $P_1({\bf{r}},s)$ and
$P_2(t,s)$ associated with the Langevin equations (2) and (5) are
easily inferred \cite{fogedby2,hugh}. From the definition
$P_1({\bf{r}},s)=\linebreak\langle\delta({\bf{r}}-{\bf{r}}(s))\rangle$ and
$P_2(t,s)=\langle\delta(t-t(s))\rangle$,
the Eqs. (1) and (4), and averaging according to the power
law noises in Eqs. (7) and (8), we deduce the scaling forms,
\cite{bouch1}
\begin{eqnarray}
P_1({\bf{r}},s)=\int\frac{d^dk}{(2\pi)^d}
e^{i{\bf{k}}\cdot{\bf{r}}-k^\mu s}=
s^{-d/\mu}G_1\left(\frac{r}{s^{1/\mu}}\right),
\end{eqnarray}
\begin{eqnarray}
P_2(t,s)=\int\frac{d\omega}{(2\pi)}
e^{-i\omega t-(-i\omega)^\nu s}=
s^{-1/\nu}G_2\left(\frac{t}{s^{1/\nu}}\right).
\end{eqnarray}
By choosing a suitable scale for $t$ and $s$ we have fixed the coefficients
of $k^\mu$ and $(-i\omega)^\nu$ to be unity.

The scaling exponents $\mu$ and $\nu$ depend on the step index
$f$ and time index $g$, respectively, characterizing the power law step size
and waiting time distributions in Eqs. (7) and (8). For
$f>2$, i.e., the case of a finite mean square step, $\mu$
locks onto the value $2$ and the scaling function $G_1$
takes the Gaussian form characteristic of ordinary Brownian walk,
$G_1(x)=\exp{(-x^2)}$. This is a consequence of the central
limit theorem which here leads to universal behaviour and
defines the universality class of Brownian motion. For $f<2$
the scaling exponent $\mu=f$, and the scaling function
$G_1$ can only be given explicitly in terms of known
functions for $\mu=1$ and $\mu=1/2$ (the Cauchy and Smirnov
distributions, respectively, \cite{shle2}). It is, however, easy
to show that $G_1\rightarrow const.$ for $x\rightarrow 0$ and
$G_1\rightarrow 0$ for $x\rightarrow\infty$. From the
distribution in Eq.~(10) we infer the scaling form for the
mean square displacement of the walker in terms of the
path variable $s$,
\begin{eqnarray}
\langle r^2(s)\rangle=\int P_1({\bf{r}},s)r^2d^dr\propto s^{2/\mu}.
\end{eqnarray}
A similar discussion applies to $P_2(t,s)$. For $g>1$,
i.e., the case of a finite first moment, corresponding to a
well defined hopping rate, the scaling exponent $\nu$
locks onto $1$ and the scaling function $G_2(x)=\delta(1-x)$.
This is again a consequence of the central limit theorem
which leads to a universal time behaviour, i.e., a
hopping rate. For $g<1$ the scaling exponent $\nu$ locks onto
$g$, i.e., $\nu = g$ and we obtain a non trivial scaling
behaviour. For the mean square displacement in terms
of the path variable we have,
\begin{eqnarray}
\langle t^2(s)\rangle=\int P_2(t,s)t^2dt\propto s^{2/\nu}.
\end{eqnarray}
In Fig. 2 we have depicted the scaling exponents
$\mu$ and $\nu$ as functions of the step index $f$ and
the waiting time index $g$, respectively.

A simple heuristic argument \cite{bouch1} using Eq. (13)
to infer that $t$ scales like $s^{1/\nu}$ and
eliminating $s$ in Eq. (12) yields the scaling relation,
\begin{eqnarray}
\langle r^2(t)\rangle\propto t^{2\nu/\mu},
\end{eqnarray}
and according to Eq. (9) the dynamic exponent,
\begin{eqnarray}
z=\frac{\mu}{\nu}.
\end{eqnarray}
As we shall discuss in section V this is indeed the
correct result following from the scaling law for $P({\bf{r}},t)$,
\begin{eqnarray}
P({\bf{r}},t)\equiv t^{-\frac{d\nu}{\mu}}G\left(r/t^\frac{\nu}{\mu}\right).
\end{eqnarray}
We conclude this section by writing down the Fokker-Planck
equations following from Eqs. (10) and (11). Introducing
the `fractional' non local differential operators,
\begin{eqnarray}
\nabla^\mu=-\int\frac{d^dk}{(2\pi)^d}e^{i{\bf{k}}\cdot{\bf{r}}}k^\mu,
\end{eqnarray}
\begin{eqnarray}
D^\nu=-\int\frac{d\omega}{2\pi}e^{-i\omega t}(-i\omega)^\nu,
\end{eqnarray}
reflecting the long range L\'{e}vy steps and waiting
times, we have
\begin{eqnarray}
\frac{\partial P_1({\bf{r}},s)}{\partial s}=\nabla^\mu P_1({\bf{r}},s)
\end{eqnarray}
\begin{eqnarray}
\frac{\partial P_2(t,s)}{\partial s}=D^\nu P_2(t,s).
\end{eqnarray}
In the Brownian case $\mu=2$ and $\nu=1$ and $\nabla^\mu$
reduces to the usual Laplace operator $\Delta$ describing ordinary diffusion,
whereas $D^\nu$ becomes the first order differential operator
$-\partial/\partial t$.

In the presence of a force field ${\bf{F}}({\bf{r}})$ we have, correspondingly,
\cite{fogedby2}
\begin{eqnarray}
\frac{\partial P_1({\bf{r}},s)}{\partial s}=
-\nabla({\bf{F}}({\bf{r}})P_1({\bf{r}},s))+\nabla^\mu P_1({\bf{r}},s).
\end{eqnarray}
Here the first term on the right hand side of Eq. (21) is the usual
drift term due to the motion of the walker in the force field.

\section{The distribution $P({\bf{\lowercase{r}}},{\lowercase {t}})$}

In order to calculate the probability distribution for the
walker as a function of the physical time $t$ we must
eliminate the path variable $s$. In other words, we have to
derive the distribution $P_3(s,t)$ since it then follows
that $P({\bf{r}},t)$ is given by the relationship,
\begin{eqnarray}
P({\bf{r}},t)=\int_o^\infty ds P_1({\bf{r}},s)P_3(s,t),
\end{eqnarray}
due to the fact that the
probability of the walker arriving at ${\bf{r}}$ in time
$t$ equals the probability of being at $s$ on the path
at time $t$ multiplied by the probability of being at
position ${\bf{r}}$ for this path length $s$, summed over all
path lengths.

In order to derive $P_3(s,t)$ we use the general expression
for $P({\bf{r}},t)$ for arbitrary step and waiting time distributions
given by Montroll and Weiss \cite{montroll}. In Fourier space we have
\begin{eqnarray}
P({\bf{k}},\omega)=(-i\omega)^{-1}\frac{1-w(\omega)}
{1-w(\omega)\pi({\bf{k}})},
\end{eqnarray}
where $\pi({\bf{k}})$ and $w(\omega)$ are the Fourier transforms of
the step and waiting time distributions,
i.e.,
\begin{eqnarray}
\pi({\bf{k}})=\int e^{-i{\bf{k}}\cdot\eta}\pi({\mbox{\boldmath$\eta$}})d^d\eta
\end{eqnarray}
\begin{eqnarray}
w(\omega)=\int e^{i\omega\tau}w(\tau)d\tau.
\end{eqnarray}
We notice that since $w(\tau)=0$ for $\tau<0$ due to causality Eq. (25)
reduces to a Laplace transform and $\omega$ is defined along
a contour parallel with the x-axis in the upper half complex
$\omega$ plane.

For the power law distributions in Eqs. (7) and (8) we obtain
in particular $\pi({\bf{k}})=1-k^\mu$ for small $k$ and
$w(\omega)=1-(-i\omega)^\nu$ for small $\omega$, i.e., to
leading order in $\omega$ and ${\bf{k}}$, the distribution
\begin{eqnarray}
P({\bf{k}},\omega) = \frac{(-i\omega)^{\nu-1}}
{(-i\omega)^\nu+k^\mu}.
\end{eqnarray}
Inserting Eq. (10) for $P_1({\bf{r}},s)$ in Eq. (22) and requiring
that $P({\bf{r}},t)$ is the Fourier transform of Eq. (26) it is easy
to demonstrate that
$P_3(s,t)$ is given by \cite{bouch1}
\begin{eqnarray}
P_3(s,t)=\int\frac{d\omega}{2\pi}
e^{-i\omega t}(-i\omega)^{\nu-1}e^{-(-i\omega)^\nu s}.
\end{eqnarray}
We have not found a simple physical argument leading
to Eq. (27), i.e., the `inversion' of Eq. (11), but notice
that $\partial P_3/\partial t=\partial P_2/\partial s$.

\section{Scaling analysis}
In the absence of the force field, i.e., for ${\bf{F}} =0$, it is
an easy task to carry out a scaling analysis. The results
are most easily deduced from Eq. (26). We have
\begin{eqnarray}
P({\bf{r}},t)=\int\frac{d\omega}{2\pi}\frac{d^dk}{(2\pi)^d}
e^{-i\omega t+i{\bf{k}}{\bf{r}}}\frac{(-i\omega)^{\nu-1}}
{(-i\omega)^\nu+k^\mu},
\end{eqnarray}
leading to the scaling form in Eq. (16) with scaling function
\begin{eqnarray}
G(x)=x^{-d}\int\frac{d\omega}{2\pi}\frac{d^dk}{(2\pi)^d}
\frac{e^{-i\omega+i{\bf{k}}}}
{(-i\omega)[1+x^{-\mu}k^\mu/(-i\omega)^\nu]}.
\end{eqnarray}
The mean square displacement is given by
\begin{eqnarray}
\langle r^2(t)\rangle=\int d^drP({\bf{r}},t)r^2\propto t^{2/z}
\end{eqnarray}
and we find the dynamic exponent $z=\mu/\nu$ in
agreement with the heuristic result in
Eq. (15).

Along the line $\mu=2\nu$ we have $z=2$ and CTLF
have the same scaling characteristics
as ordinary Brownian motion. The superdiffusive behaviour
induced by the long range L\'{e}vy steps is precisely
balanced by the long waiting times. We notice, however, that
the scaling function $G$ depends on the scaling index $\mu=2\nu$;
only in the case $\mu=2$ and $\nu=1$ do we obtain the
Gaussian distribution.

For $\mu>2\nu$ we have $z>2$, the L\'{e}vy flights prevail, and we obtain
superdiffusive behaviour; correspondingly, $\mu<2\nu$ implies $z<2$, the
long waiting times dominate, and we have subdiffusive behaviour.

In Fig. 3 we have shown the different universality classes
for CTLF. For $z=2$ ($\mu=2\nu$) we have the universality class
of ordinary Brownian motion. For $\mu>2\nu$ we obtain the
universality class (or classes) of anomalous superdiffusion
with an exponent $z$ depending continuously on the ratio of
the microscopic exponents $f$ and $g$. Similarly, for
$\mu<2\nu$ we have the universality class of anomalous subdiffusion.

\section{Discussion and conclusion}
In the present paper we have discussed the combined effects of
an algebraic waiting time distribution and an algebraic L\'{e}vy
type step distribution on the motion of a random walker.
In order to include external force fields we have formulated this
analysis in terms of a set of coupled Langevin equations. In the
absence of force fields a simple scaling analysis shows that the
dynamic exponent $z$ characterizing the long time behaviour of
the mean square displacement is given by the ratio
$z=\mu/\nu$, where the scaling exponents $\mu$ and $\nu$
are related to the microscipic step and waiting time exponents
($\mu=f$ for $f<2$ and $\nu=g$ for $g<1$). This dependence defines
three universality classes: 1) Normal diffusive behaviour for
$\mu=2\nu$, 2) Anomalous superdiffusion for $\mu>2\nu$,
3) Anomalous subdiffusion for $\mu<2\nu$.

In the presence of a quenched (time independent) Gaussian random force field
and a well defined hopping rate, i.e., for $\nu=1$, we have recently shown
\cite{fogedby2} that $z$ locks onto $\mu$ for $f<2$. It is now easy to
generalize this result to the case of a non trivial waiting time
distribution with $\nu<1$. Since the force field is independent of $t$
we can directly apply the discussion in ref. \cite{fogedby2} to Eq. (3)
treating the path variable $s$ as the effective time. We conclude
that in this case $z$ is also given by $\mu/\nu$ for weak quenched
disorder. The analysis in ref. \cite{fogedby2}, leading to
the
critical dimension $d_c=2\mu-2$ below which a non trivial force correlation
fixed point emerges can also be carried over to the present case.
{}From a physical point of view it is clear that a nontrivial
waiting time distribution has no effect since the quenched force field
acting at position $r$ simply `waits' till the walker arrives.
However, in the case of a time dependent random force field
the waiting time distribution will interfere with the temporal
force correlations and we have to treat the coupled Langevin
equations in order to eliminate the intermediate path variable $s$.
This interesting case will be considered in a
forthcoming publication \cite{fogedby3}.

\acknowledgements
The author wishes to thank K. B.
Lauritsen and A. Svane for helpful discussions.
This work was supported by the Danish Research Council,
grant no. 11-9001.

\begin{figure}
\setlength{\unitlength}{1cm}
\begin{picture}(7,6)(-1,-1)
\thicklines
\put(0,0){\line(1,2){1}}
\put(1,2){\line(1,-2){0.5}}
\put(1.5,1.0){\line(1,2){1}}
\put(2.5,3.0){\line(1,-1){0.5}}
\put(3.0,2.5){\line(1,-3){0.5}}
\put(3.5,1.0){\vector(1,2){1}}
\put(2.2, 2.0){\makebox(0,0){$s$}}
\end{picture}
\caption
{
Plot of a particular random walk parametrized by
the arc length $s$
}
\setlength{\unitlength}{1cm}
\begin{picture}(6,5)(-1,-1)
\put(0,0){\line(0,1){3}}
\put(0,0){\line(1,0){4}}
\put(2,0){\line(0,1){0.2}}
\put(1,0){\line(0,1){0.2}}
\put(0,2){\line(1,0){0.2}}
\put(0,1){\line(1,0){0.2}}
\thicklines
\put(0,0){\line(1,1){2}}
\put(2,2){\line(1,0){2}}
\put(4,-0.5){\makebox(0,0){f}}
\put(-0.5,3){\makebox(0,0){$\mu$}}
\put(3,3){\makebox(0,0){(a)}}
\put(2,-0.5){\makebox(0,0){2}}
\put(1,-0.5){\makebox(0,0){1}}
\put(0,-0.5){\makebox(0,0){0}}
\put(-0.5,2){\makebox(0,0){2}}
\put(-0.5,1){\makebox(0,0){1}}
\put(-0.5,0){\makebox(0,0){0}}
\end{picture}
\begin{picture}(6,5)(-1,-1)
\setlength{\unitlength}{1cm}
\put(0,0){\line(0,1){3}}
\put(0,0){\line(1,0){4}}
\put(1,0){\line(0,1){0.2}}
\put(0,1){\line(1,0){0.2}}
\thicklines
\put(0,0){\line(1,1){1}}
\put(1,1){\line(1,0){1.5}}
\put(4,-0.5){\makebox(0,0){g}}
\put(-0.5,3){\makebox(0,0){$\nu$}}
\put(3,3){\makebox(0,0){(b)}}
\put(1,-0.5){\makebox(0,0){1}}
\put(0,-0.5){\makebox(0,0){0}}
\put(-0.5,1){\makebox(0,0){1}}
\put(-0.5,0){\makebox(0,0){0}}
\end{picture}
\caption{
In a) we show $\mu$ as a function of $f$. For $f>2$ the exponent
$\mu = 2$ and we have ordinary Brownian walk;
for $f<2$ the exponent $\mu = f$ and we have L\'{e}vy flights,
leading to anomalous superdiffusion. In b) we depict
$\nu$ as a function of $g$. For $g>1$ the exponent $\nu=1$ and we have a
well defined hopping rate for the walker; for $g<1$
we have $\nu=g$ and we obtain a subdiffusive behaviour.
}
\setlength{\unitlength}{1cm}
\begin{picture}(6,6)(-1,-1)
\put(0,0){\line(0,1){4}}
\put(0,0){\line(1,0){4}}
\put(1.5,0){\line(0,1){3}}
\put(0,3){\line(1,0){1.5}}
\put(0,1.5){\line(1,0){0.2}}
\thicklines
\put(0,0){\line(1,2){1.5}}
\put(4,-0.5){\makebox(0,0){$\nu$}}
\put(-0.5,4){\makebox(0,0){$\mu$}}
\put(0.5,2.5){\makebox(0,0){{\small super}}}
\put(1,0.5){\makebox(0,0){{\small sub}}}
\put(1.5,-0.5){\makebox(0,0){1}}
\put(0,-0.5){\makebox(0,0){0}}
\put(-0.5,3){\makebox(0,0){2}}
\put(-0.5,1.5){\makebox(0,0){1}}
\put(-0.5,0){\makebox(0,0){0}}
\end{picture}
\caption{
Plot of $\mu$ versus $\nu$. Along the solid line $\mu=2\nu$ we have
the universality class of ordinary Brownian motion with $z=2$.
In the region $\mu>2\nu$ we have the universality class of
anomalous superdiffusion; in the region $\mu<2\nu$ the class
of anomalous subdiffusion.
}
\end{figure}

\begin{thebibliography}{99}
\bibitem{rw}	We use the expression `random walk' in the
general sense of a walk generated by arbitrary step and
waiting time distributions.
\bibitem{bouch1} J.-P. Bouchaud and A. Georges,
		Phys. Rep. {\bf 195}, 127 (1990)
\bibitem{scher1}H. Scher and M. Lax,
		Phys. Rev. {\bf B7}, 4491 (1973)
\bibitem{scher2}H. Scher and E.W. Montroll,
		Phys. Rev. {\bf B12}, 2455 (1975)
\bibitem{shle1} M.F. Shlesinger,
		J. Stat. Phys. {\bf 10}, 421 (1974)
\bibitem{klaf1} J. Klafter and R. Silbey,
		Phys. Rev. Lett. {\bf 44}, 56 (1980)
\bibitem{blumen1} A. Blumen, J. Klafter, and G. Zumofen,
		Phys. Rev. {\bf B27}, 3429 (1983)
\bibitem{man}	P. Manneville,
		J. Physique (Paris) {\bf 41}, 1235 (1980)
\bibitem{gei}   T. Geisel and S. Thomae,
		Phys. rev. Lett. {\bf 52}, 1936 (1984)
\bibitem{pom}	Y. Pomeau, A. Pumir, and W. Young,
		Phys. Fluids {\bf A1}, 462 (1989)
\bibitem{car}   O. Cardoso and P. Tabeling,
		Europhys. Lett. {\bf 7}, 225 (1988)
\bibitem{fogedby1} H.C. Fogedby, T. Bohr, and H.J. Jensen,
		J. Stat. Phys. {\bf 66}, 583 (1992)
\bibitem{feller} W. Feller, {\em An Introduction to Probability
                Theory and its Applications} (Wiley, N.Y. 1971)
\bibitem{klaf2} J. Klafter, A. Blumen, G. Zumofen, and
                M.F. Shlesinger, Physica A {\bf 168}, 637 (1990)
\bibitem{blumen2} A. Blumen, G. Zumofen, J. Klafter,
                Phys. Rev. {\bf A40}, 3964 (1989)
\bibitem{ott} A. Ott, J.-P. Bouchaud, D. Langevin, and W. Urbach,
                Phys. Rev. Lett. {\bf 65}, 2201 (1990)
\bibitem{shle2} M.F. Shlesinger, J. Klafter, and Y.M. Wong,
		J. Stat. Phys. {\bf 27}, 499 (1982)
\bibitem{weiss}	G.H. Weiss and R.J. Rubin,
		Adv. in Chem. Phys. {\bf 52}, 363 (1983)
\bibitem{fogedby2} H.C. Fogedby, IFA Report No. 94/02 (cond-mat/9401007)
\bibitem{hugh} B.D. Hughes, M.F. Shlesinger, and E.W. Montroll,
                Proc. Natl. Acad. Sci. USA {\bf 78}, 3287 (1981)
\bibitem{montroll} E.W. Montroll and G.H. Weiss,
		J. Math. Phys. {\bf 6}, 167 (1965)
\bibitem{fogedby3} H.C. Fogedby (unpublished)
\end{thebibliography}
\end{document}